\documentstyle[preprint,pra,aps]{revtex}

 \input epsf

\begin{document}

\vspace{1cm}

\begin{center}
  {\bf
Interference effects in electronic transport through metallic
single-wall carbon nanotubes }

\vspace{0.5cm} S.~Krompiewski$^1$, J.~Martinek$^{2,3,1}$, and
J.~Barna{\'s}$^{1,4}$

\vspace{0.3cm}
 $^1$Institute of Molecular Physics, Polish Academy
of Sciences, \\ ul. M. Smoluchowskiego 17, 60-179 Pozna\'n,
Poland\\
 $^2$ Institute for Materials Research, Tohoku University,
Sendai 980-8577, Japan\\
 $^3$Institut f\"{u}r Theoretische Festk\"{o}perphysik,
 Universit\"{a}t Karlsruhe, 76128 Karlsruhe, Germany \\
 $^4$ Adam Mickiewicz University,
Department of Physics, \\ ul. Umultowska 85, 61-614 Pozna{\'n},
Poland

\end{center}

\vspace{0.5cm}
 {\small  In a recent paper Liang {\it et al.}
[Nature {\bf 411}, 665 (2001)] showed experimentally, that
metallic nanotubes, strongly coupled to external electrodes, may
act as coherent molecular waveguides for electronic transport. The
experimental results were supported by theoretical analysis based
on the scattering matrix approach. In this paper we analyze
theoretically this problem using a real-space approach, which
makes it possible to control quality of interface contacts.
Electronic structure of the nanotube is taken into account within
the tight-binding model. External electrodes and the central part
(sample) are assumed to be made of carbon nanotubes, while the
contacts between electrodes and the sample are modeled by
appropriate on-site (diagonal) and hopping (off-diagonal)
parameters. Conductance is calculated by the Green function
technique combined with the Landauer formalism. In the plots
displaying conductance {\it vs.} bias and gate voltages, we have
found typical diamond structure patterns, similar to those
observed experimentally. In certain cases, however, we have found
new features in the patterns, like a double-diamond sub-structure.

\vspace{0.5cm}


\newpage

There has been o lot of interest of physicists and technologists
in carbon nanotubes since they were discovered a decade ago
\cite{Iijima}. This is due to their outstanding mechanical and
electrical properties, not only fascinating from the fundamental
point of view but also very promising for practical applications
in molecular electronics -- for instance as field emitters for
flat-panel displays, current rectifiers, single-electron
transistors, chemical sensors (important for environment
protection) or hydrogen storage media (to power electric
engines)~\cite{book,dekker,phys}. Carbon nanotubes display a rich
spectrum of physical properties which make them  attractive for
both experimental and theoretical studies. First of all, their
electrical properties depend critically on the way a graphene
(honey-comb) sheet is rolled up, i.e., on its chirality. The
chirality is defined by two integer numbers ($n, m$), which
determine the wrapping vector on the graphene sheet\cite{dekker}
(the length of which is equal to the circumference of the
nanotube). The basic orientations of nanotubes are: (i) the
so-called "armchair" with $n=m$, which is metallic, and (ii) the
"zigzag" one ($n \ne 0, m=0$). The nanotubes are metallic when
$n-m=3 \times integer $, and semiconducting otherwise.  They may
span in length from a few nanometers up to hundreds $\mu$m, with a
typical diameter of the order of a few nm, or more -- especially
in the case of multi-wall nanotubes.

Electronic transport through a nanotube depends on the contacts
with electrodes. If, e.g., electrodes are weakly coupled to a
single-wall carbon nanotubes (SWCNTs), one observes Coulomb
blockade effects. Additionally, some evidences of the Luttinger
liquid behavior were found in that limit\cite{set}. In the
opposite case (strong coupling), one expects armchair nanotubes to
approach the ballistic limit with the conductance up to $4 e^2 /h
=4 (6.45 k\Omega)^{-1}$ (twice the conductance quantum), where the
factor 4 is due to spin degeneracy and the two bands crossing the
Fermi energy. In the intermediate region, the Kondo resonance was
observed~\cite{Nygard}.

In a recent paper Liang {\it et al} \onlinecite{Liang}
investigated electronic transport through metallic carbon
nanotubes strongly coupled to external electrodes, and observed
features in the conductance, which were typical of interference
effects. The nanotubes behaved as coherent molecular waveguides
for electronic transport. When the conductance was plotted in a
gray scale against the bias and gate voltages, a characteristic
diamond structure was revealed. Experimental observations were
supported  by theoretical analysis based on the scattering matrix.

In this paper, on the other hand, we consider this problem in more
detail, using a real-space approach.  Such an approach
intrinsically takes into account the interference phenomena due to
multiple reflections from the boundaries. We show, that in some
cases the diamond structure can be split and double diamond
patterns may appear. For simplicity, we assume that the external
electrodes (called also leads in the following) and the sample
(central part) are made of similar SWCNTs. Thus, the whole model
system consists of semi-infinite left and right leads and the
central part of finite length.  The interfaces (contacts)  between
the electrodes and the sample are considered as defects and are
modeled by appropriate on-site and hopping parameters, which
differ from the reference bulk values.
In the experimental situation of Ref.\onlinecite{Liang} metallic
Au/Cr electrodes were used. However, the key features of
electronic transport studied there were due to the central part
(sample) and  its contacts to the electrodes. The electrodes themselves were
thick enough to be considered as ideal leads. In our case such
ideal leads are simulated by defect-free SWCNTs. In both cases the
contribution of electrodes to the total resistance is negligible,
and therefore one may expect that our results  describe properly
the experimental observations.

We adopt here the standard single-band tight-binding Hamiltonian
to describe the carbon nanotubes with one $\pi$-electron per
carbon atom
\begin{equation} \label{hamiltonian}
\hat H = \sum \limits_{I ,J} T_{I,J}\vert I><J \vert +\sum
\limits_{I} D_I \vert I><I \vert,
\end{equation}
where $\vert I>$ and $ \vert J> $ stand for orbitals related to
the nanotube unit cells (or principle rings, in analogy to
principle layers exploited in layered structure calculations
\cite{umerski}). It is a very convenient way of ordering
structural units which enables one to express the Hamiltonian in a
block-tridiagonal form. The off-diagonal matrices are denoted by
$T$, whereas the on-diagonal ones by $D$. The rank of the blocks
is equal to the product of the number of rings within the unit
cell times the number of circumferential carbon atoms. In the case
of ($n,n$) armchair structures there are two carbon rings in the
unit cell (bi-ring), each having $2n$ atoms, thus the product
gives $4n$ atoms. It is noteworthy to stress in this context that
within the ballistic regime, when there are neither impurities nor
thermal excitations in the system, the conductance of armchair
SWCNTs does not depend on their circumference, since it usually
remains fully determined by just two energy bands which cross the
Fermi energy. This is the case also in our situation, because for
bias voltages $V$ of our interest higher energy bands do not enter
the 'transport window' ($E_F \pm eV/2$) around the Fermi energy,
except for huge nonotube diameters corresponding to $n>420$.

The number of diagonal blocks in the Hamiltonian is equal to the
length of a nanotube expressed in lattice constant units ($a=2.49
\AA $). This is because for armchair structures there is just one
bi-ring per lattice constant (see Fig.~1). Altogether, for
armchair nanotubes of lengths $1000a - 2000a$, this gives a rank
of the whole Hamiltonian up to 8000$n$, i.e 24000 for $n=3$
considered hereafter. Fortunately, owing to the recursive Green
function method this problem is still tractable. The recursive
algorithm proceeds in the following steps:

({\it i}) The left and right surface retarded Green functions
($g_{0,0}^L$, $g_{N+1,N+1}^R$) are calculated by means of the
standard iteration technique. Incidentally, any techniques which
require inversion of the $T$ matrices are useless in our case
because for the nearest neighbor hopping approximation the $T$
matrices  are singular.

({\it ii}) Going from $I=N$ down to $I=1$, the remaining Green
functions $g^R$ are found from

\begin{equation} \label{gR}
 g_{I,I}^R=[E-D_I- T_{I,I+1} g_{I+1,I+1}^R T_{I+1,I}]^{-1}.
\end{equation}

({\it iii}) The total Green function $G_{1,1}$ can then be found as follows

\begin{equation} \label{G11}
 G_{1,1}=[E-D_1- T_{1,0} g_{0,0}^L T_{0,1}-
T_{1,2} g_{2,2}^R T_{2,1}]^{-1}.
\end{equation}

({\it iv}) The above specified matrix elements of $G$ are
sufficient to express conductivity via the transmission matrix
${\cal T}(E)$ given by (cf.~\cite{Lake})

\begin{equation} \label{T}
{\cal T}(E)=Tr(\Gamma_{1,1}^L (A_{1,1}-G_{1,1} \Gamma_{1,1}^L
G_{1,1}^{\dagger}))
\end{equation}
with $$ \Sigma^L=T_{1,0}g_{0,0}^L T_{0,1} \hspace{.1cm},
\hspace{1cm} \Sigma^R=T_{N,N+1}g_{N+1,N+1}^R T_{N+1,N}, $$ $$
\Gamma^{L,R}=i [\Sigma^{L,R}-(\Sigma^{L,R})^{\dagger}]
\hspace{.1cm} , \hspace{.5cm} A=i (G-G^{\dagger}) $$
 and the trace
taken over all orbitals in the unit cell. It is worth to
point out that the transmission matrix ${\cal T}(E)$ includes
all the relevant interference effects.

Restricting ourselves to phase-coherent transmission, and assuming
a constant potential within the whole sample (central electrode),
we calculate the current from the following formula
\begin{equation} \label{IV}
I(V) = \frac{2e}{h} \int_{-\infty}^{\infty}[f(E-\frac{eV}{2})-
f(E+\frac{eV}{2})]{\cal T}(E, \Phi) \,dE,
\end{equation}
where $f(E)$ is the zero temperature Fermi-Dirac distribution
function, and $\Phi$ is the additional electrostatic energy in the
central electrode generated by the gate voltage $V_g$ (we assume
that only the central part is subject to the gate voltage). Hence,
the differential conductance reads
\begin{equation} \label{G}
  \frac{\partial I(V)}{\partial V}=\frac{e^2}{h} [{\cal T}(E_F+\frac{eV}{2},\Phi)+{\cal T}(E_F-\frac{eV}{2},\Phi)].
\end{equation}

Following Liang et al. \cite{Liang} we assume that for symmetric
contacts the access charge $Q$ in the central electrode scales
with $V_g$ as

\begin{equation} \label{Q}
Q=L C V_g,
\end{equation}
where $L=Na$. Equation (\ref{Q}), when combined with the estimate
of $Q$ in terms of the density of states $Q=- ({8L}/{h v_F})\Phi
$, leads to the universal relation between the gate voltage and
the average potential energy $\Phi$
\begin{equation} \label{Phi}
 \Phi=- \frac{C}{4} \frac{h v_F}{2} V_g ,
\end{equation}
where $C$ stands for the capacitance per unit length of the sample
($C=20 V^{-1} \mu$m$^{-1}$ according to \cite{Liang}), and $v_F$
is the Fermi velocity related to the slope of the energy level
intersecting the Fermi energy ($ hv_F/2=$1.67meV$\mu$m).

Equations (\ref{T}) and (\ref{G}) form the basis for our analysis
of the differential conductance of armchair nanotubes. Our
approach -- as a real-space method -- makes it possible to model
arbitrarily the sample and the contacts, while keeping the
external leads ideal. The contacts will be considered as defects
in the otherwise perfect structure (disregarding the gate
voltage). So far, tight-binding studies of nanotubes have focused
on diagonal disorder and topological defects related to bending
and twisting of nanotubes
\cite{Chico,Anantram,Kostyrko,Orlikowski,Rochefort2}. No
systematic studies taking into account both diagonal and
off-diagonal disorder in the presence of charging effects in the
sample have been carried out yet, to our knowledge. The recursive
algorithm presented above enables us to handle very complex
structures, with arbitrary distribution of the hopping integrals
and on-site potentials in the sample. However, in our case we vary
only the parameters of the contacts, i.e., the contact nearest
neighbor hopping integral $t_c$ (which enters the left and right
corners of the matrices $T_{0,1}$ and $T_{N,N+1}$, respectively)
and the on-site potential energy $\epsilon$ (that comes to the
respective upper and lower semi-diagonals of $D_1$ and $D_N$).
Thus, we have just two fitting parameters related to the contacts,
while all the nearest neighbor hoppings are set equal to
$t_{i,j}=-1$, and all the atomic on-site energies in the leads are
set equal to zero, whereas in the sample the on-site energies are
increased by $\Phi$ due to charging caused by the gate voltage.

Conductance of a ballistic conductor oscillates as a function of
length, but it does not decay (non-ohmic behavior) \cite{Datta}.
The key role played by the length consists in introducing an
energy scale determined by the separation of energy levels. This
is of crucial importance in the Coulomb blockade and in the Kondo
regimes. It is also relevant in the strong coupling limit, and
determines the periods in conductance oscillations with $V$ and
$V_g$. Another noteworthy point is that armchair SWCNTs show a
conductance peak with a period $3a$, as their length varies. The
physics standing behind this is quite simple and is due to the
existence of zero energy eigen-value which coincides with the
Fermi energy for such a length (in the ideal
case, allowed values for a quantized wave vector include $k=2
\pi/(3a)$ then \cite{Orlikowski}). 
We will see that there is a reminiscence of this
behavior even when SWCNTs have modified interface parameters and
are coupled to external electrodes. Our approach which makes also
possible to sample SWCNT lengths equal to odd multiples of $a/2$
(single inter-ring spacing) shows that a fundamental quasi-period
is $3 \cdot a/2$ rather then $3a$, confirming thereby an earlier
observation of Ref.~\onlinecite{Rochefort1} in the context of
energy spectra of isolated SWCNT segments. To be more specific, we
present our results for nanotubes 884$a$ and 2129$a$ long (220~nm
and 530~nm, as in Ref. \cite{Liang}), which contain an even number
of rings (integer number of elementary bi-rings), and also for
nanotube lengths incommensurate with the lattice constant $a$
(883.5$a$ and 2128.5$a$).

Our main results are presented in Figs 2 and 3 in a form of
two-dimensional gray-scale density plots representing the
conductance as a function of the bias and gate voltages. In order
to convert $V$ and $V_g$ to Volts they should be multiplied by
$|t_{i,j}|$ in eV (typically 2.7 \cite{White}). We have chosen the
parameters for the contact hopping integral across the interfaces
and the on-site potential at the inner interface rings of carbon
atoms so as to obtain variations of the conductance values similar
to those of Ref.\onlinecite{Liang}. We have found that this would
not be possible with only one fitting parameter, no matter whether
$t_c$ or $\epsilon$. The parameters we use are: $ t_c=-0.73 $ and
$ \epsilon=0.07 $ for Fig.~2, and $t_c=-0.6$ and $\epsilon=0.35$
for Fig.~3. The results presented in Figs~2 and 3 agree with the
experimental data in terms of both the conductance values and the
periods in $V$ and $V_g$.
The large value of conductance
indicates an almost perfect ballistic transport. Moreover, the periods
in $V$ and $V_g$ are determined by the interlevel spacing as expected
in the ballistic regime, whereas in the Coulomb blockade regime,
the periods would be mainly determined by charging energy.
For the periods in the bias voltage, $\Delta
V$, we have found (for $t_{i,j}=-2.7 eV$) 16.7 mV and 6.8 mV,
which agree with the estimated values from the positive
interference condition, $ \Delta V=hv_F/eL$, giving 15.2mV and
6.78mV for the 220 nm and 530 nm SWCNTs, respectively.  The
corresponding experimental values of Ref.\onlinecite{Liang} are
13mV  and 7mV.
    Thus, the calculated interference patterns can be indeed interpreted
 in terms of  multiple electron reflections and superposition
 with the phase change $(\frac{2 \pi}{\lambda}\cdot 2L) $ corresponding to
the round trip length $2L$ and the wavelength
 $ \lambda=2 h v_F /(e \Delta V)$.
Some discrepancies concerning the phases and certain
irregularities present in experimental data are probably due to
different external electrodes used in Ref.\onlinecite{Liang} and
deviation of real systems from ideal ballistic conductors.

A detailed analysis of our results reveals new features in the
diamond structure, like for instance the double-diamond patterns,
indicated by dashed lines in Fig.~3. Such patterns originate from
reflection conditions at the interfaces and are more pronounced
when the contact hopping integral $t_c$ differs considerably from
the ideal value $t_{i,j}$, as it can be seen in Fig.3, in contrast
to Fig.2 (with smaller $|t_c-t_{i,j}|$). This is illustrated in
more detail in Fig.4. The thick solid line is there a $V=0$
cross-section of Fig.~3 and its main minima are related to the
dominant diamonds, whereas the local minima  to the faint ones.
For a nanotube $2128.5a$ in length, represented by the thin full
line, only main minima are present, implying the appearance of
just one type of diamonds. A similar scenario holds for $884a$,
and $883.5a$ long SWNTs  (dashed curves in Fig.~4), but then the
shallow dips (shoulders) of the thick dashed line are too small
for the diamond sub-structure to be visible in Fig.~2. The
appearance of the double-diamond structure may be explained in
terms of symmetry induced resonant level degeneracy. 
   Our structural models are symmetric by construction (see Fig.~1) even
   if the contact parameters differ considerably from the reference
   ones. Now, for varying SWCNT lengths the on-resonance conductance peak
   (close to $E_F$)  will be split as a result of lifting of degeneracy
   by modified boundary conditions at the interfaces. The process will
   repeat with a quasi-period $3 \cdot a/2$.

There is no evidence of the double diamond sub-structure in the
experimental data of Ref.\onlinecite{Liang}. However, our results
indicate that such a structure may be observable. The experimental
data for 530 nm long SWCNTs show clearly a superstructure imposed
on the basic diamond structure. Such a superstructure may result
from superposition of at least two different oscillation periods
in both bias and gate voltages. Since the periods depend on the
sample length, one may expect that the superstructure is due to
some additional defects inside the sample. Such defects may
produce a diamond structure with longer periods. When imposed onto
the main pattern, it can then lead to the patterns observed
experimentally.  This is a possible explanation of the
experimental results (for 530 nm SWCNT), but probably not the only
one. There might be some other physical mechanisms producing
similar effects, which have not been included in our calculations.

In conclusion, our calculations yield correct values of both the
conductance oscillation periods and the conductances themselves.
Apart from this, our results show that the diamond structure
patterns for ${\partial I(V,V_g)}/{\partial V}$ are very sensitive
to conditions at the contacts between the sample and external
leads. The off-diagonal disorder, described by a modified hopping
integral across the interfaces, acts as moderate energy barriers
(strong coupling or weak confinement limit). The periods in the
bias and gate voltages are determined by separation of the energy
levels, and are inversely proportional to the sample length. The
diagonal defects, introduced by setting extra on-site energy on
the first ring of the nanotube and on the last one, act as an
additional reflecting factor responsible for asymmetry with
respect to $V_g=0$. These two kinds of defects, only when both
taken into account, give satisfactory magnitudes of the
conductance values for various SWNT lengths. The double-diamond
paterns may be expected in the "on-resonance" case, i.e. for such
a length which makes the density of states have a peak at the
Fermi energy.

\vspace{1cm}
 {\it Acknowledgments }
  We
 thank the Pozna\'n Supercomputing and Networking Center for the
 computing time. The support through the Research Project
 5 P03B 091 20 is also acknowledged (JM).

\newpage

\listoffigures


\newpage
\begin{figure}[ht]
\vspace{1cm}
 \epsfxsize 15cm
 \epsfbox{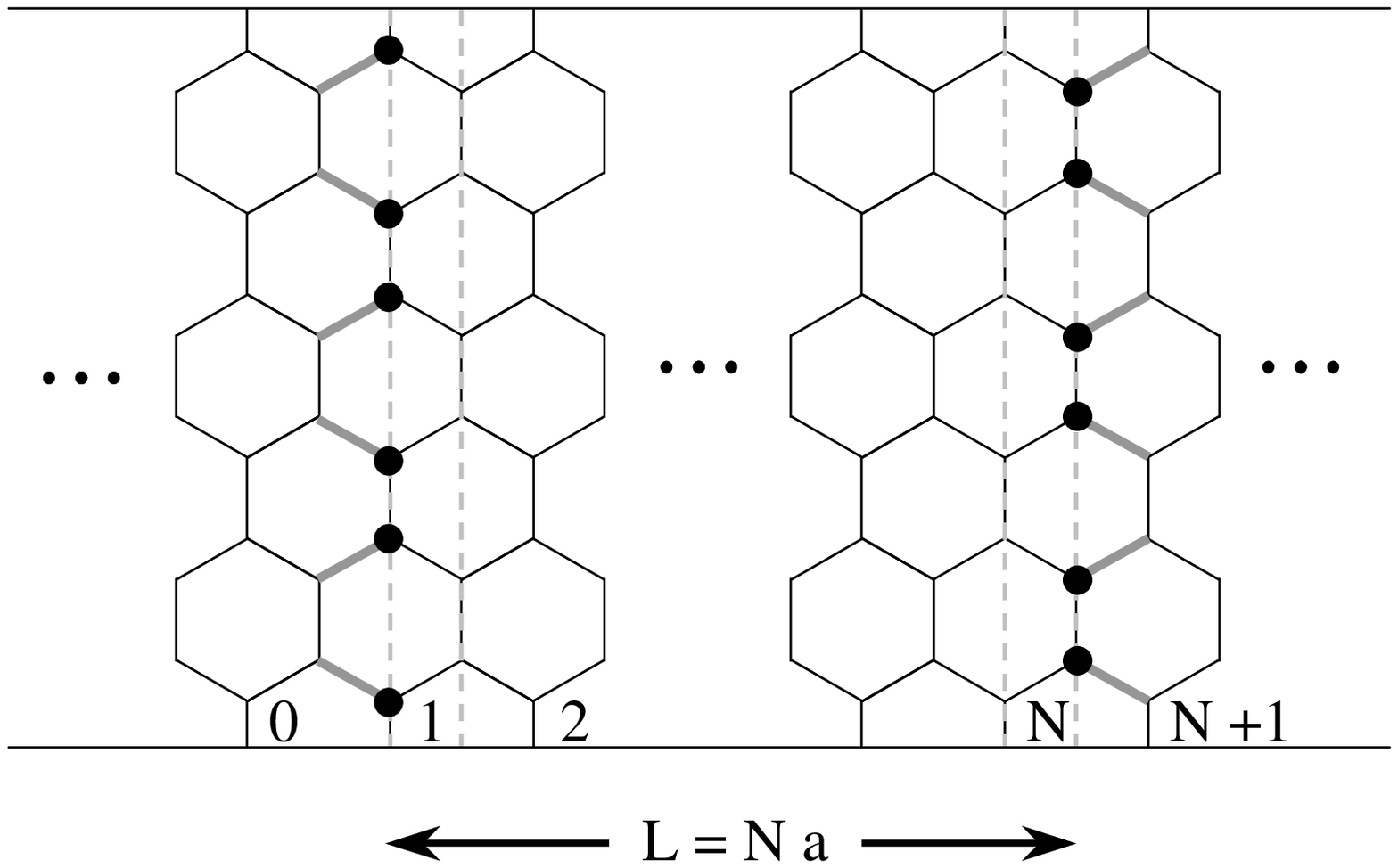}
\vspace{1cm}
 \caption{ Schematic of the device under
consideration. Both the central segment (unit cells 1,..,N) and
the electrodes are ideal single-wall carbon nanotubes (SWCNT) with
$t_{i,j}=-1$, $\epsilon=0$. Only the hopping integral across the
interfaces (heavy lines) and the on-site potential (full circles)
at the ends of the central part are modified. The figure is to be
folded up, so that the horizontal lines coincide with each other.}
\label{fig1}
\end{figure}


\newpage
\begin{figure}[ht]
 \epsfxsize 15cm
 \epsfbox{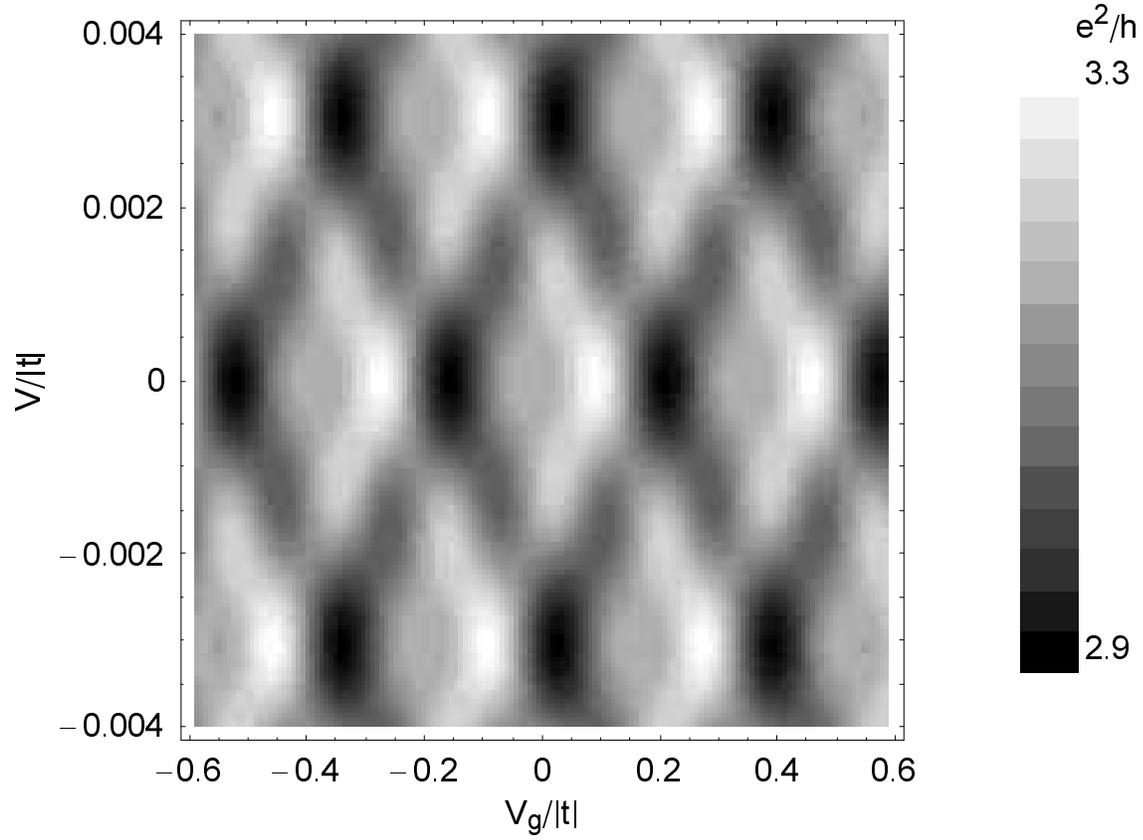}
\vspace{1cm}

\caption{ Two-dimensional ${\partial I}/{\partial V}$ plot against
bias ($V$) and gate ($V_g$) voltages, for $N=884a$ ($a=0.249 nm$),
$t_c=-0.73$ and  $\epsilon =0.07$. A conversion factor to Volts is
$ \left| t \right| $ in eV (typically 2.7).}

\label{fig2}
\end{figure}


\newpage
\begin{figure}[ht]
 \epsfxsize 15cm
 \epsfbox{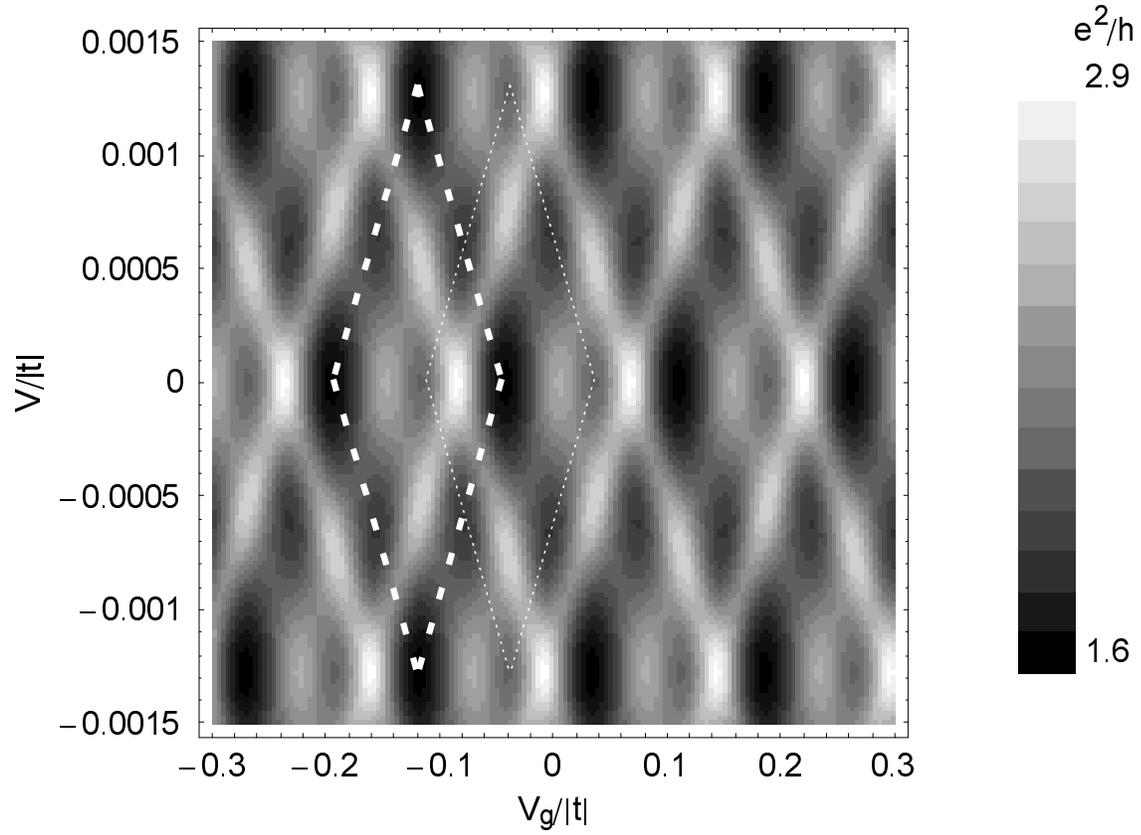}  

\vspace{1cm}

\caption{ Same as Fig.2 but for $N=2129a$,
         $t_c=-0.6$  and $\epsilon =0.35$. The dashed lines are
guides to the eye to visualize the diamond sub-structure. }
\label{fig3}
\end{figure}


\newpage

\begin{figure}[ht]
 \epsfxsize 15cm \epsfbox{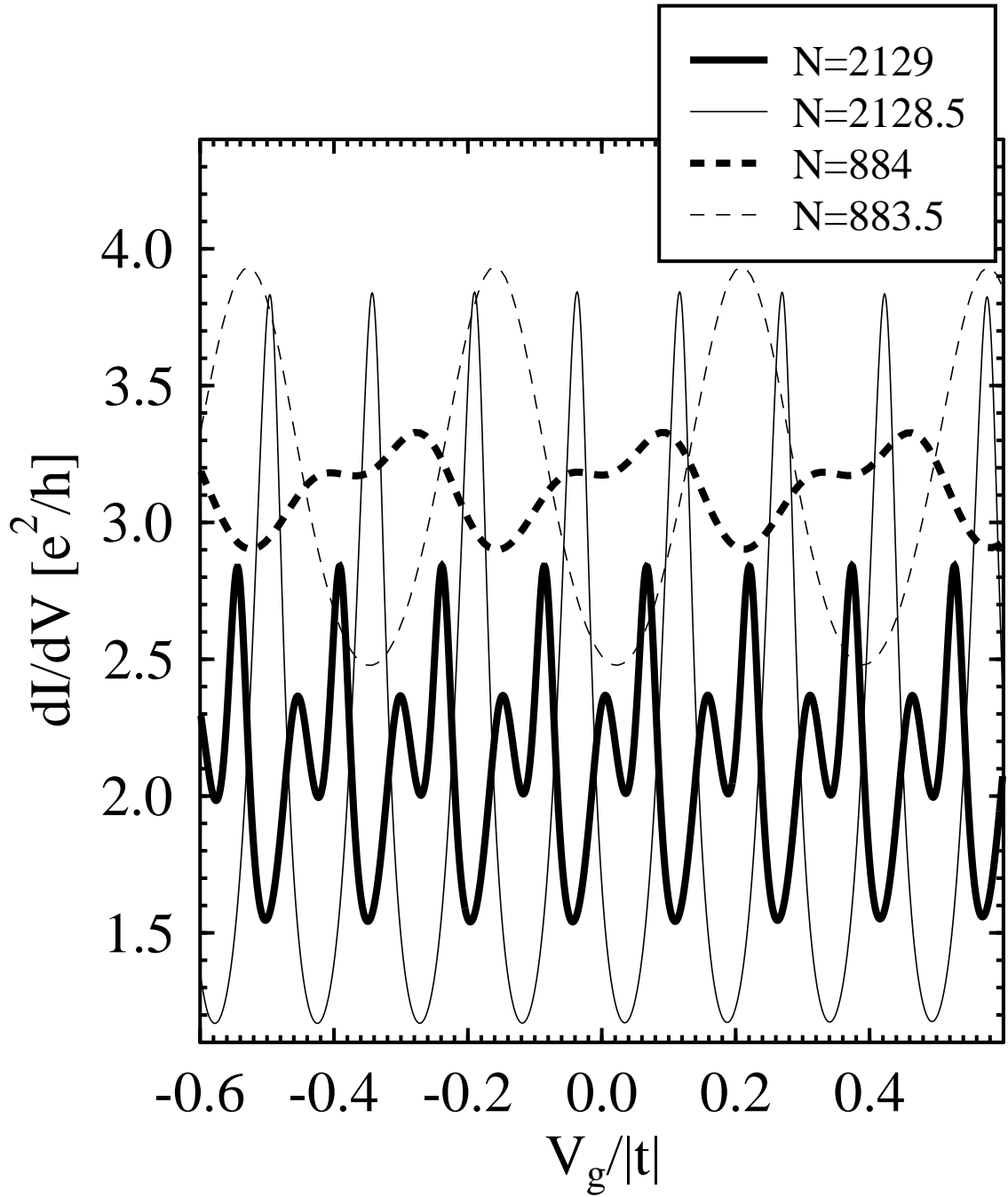}  

 \vspace{1cm}

 \caption{
${\partial I}/{\partial V}$ vs. $V_g$ at V=0 plotted for various
lengths (in lattice constant units). The diamond sub-structure
seen in Fig.3 is connected with two types of dips as those
revealed by the thick solid curve. The thick dashed line, in turn,
with just one type of pronounced minima leads to simple diamond
structure (see Fig.~2). The parameters $t_c$ and $\epsilon$ for
the dashed and full lines are as in Figs 2 and 3, respectively.}
\label{fig4}
\end{figure}

\end{document}